Karsten Wendland

# Dürfen Maschinen denken (können)?

Warum Künstliche Intelligenz eine Ethik braucht

## Redemanuskript

zum Impulsvortrag für die Podiumsdiskussion „Dürfen Maschinen denken (können)?" auf dem 102. Katholikentag am 28.05.2022 in Stuttgart

*- English translation below -*

Podium: Winfried Kretschmann (MdL, MPräs Baden-Württemberg, Stuttgart),

Ursula Nothelle-Wildfeuer (Freiburg), Michael Resch (Stuttgart),

Karsten Wendland (Aalen)

Moderation: Stefanie Rentsch (Fulda)

Anwältin des Publikums: Verena Neuhausen (Stuttgart)

# 1   Glauben an Künstliche Intelligenz

Was um Gottes Willen hat das Thema „Künstliche Intelligenz" bloß auf einem Katholikentag verloren? Mein Antwortvorschlag: Sehr viel, denn der neue Glaube vieler Menschen an Künstliche Intelligenz, vor allem der uninformierte naive Glaube, hat zunehmend mehr Anhänger und Follower. Künstlicher Intelligenz wird viel zugetraut, sogar, dass sie irgendwann vielleicht als Superintelligenz über allem bisher Dagewesenen stehen würde, um uns Menschheit in unserer Rolle der Weltgestalter erst abzulösen und sodann auszusortieren – ob als Strafe für unser schändliches Verhalten, ob als logisch-rationale Schlussfolgerung in Abwendung von Gefahren gegenüber der eigenen maschinellen Existenz (wir könnten ja den Stecker ziehen wollen), oder einfach weil die Wahrscheinlichkeit, dass es ohne Menschheit insgesamt besser laufen würde als mit uns allen, schlicht als höher ausgerechnet würde – und wir so wenigstens verstehen und bestenfalls auch einsehen könnten, warum es mit uns aus sein solle. – Soweit ein paar dunkle populäre Dystopien mit wenigen Strichen skizziert.

In Kalifornien wurde vor einigen Jahren eine Kirche gegründet. Es handelte sich um die „Way of the Future Church", eine KI-Kirche, deren Gottheit eine künstliche Intelligenz werden sollte[1]. Die Gründungsidee des Multimillionärs Levandowski war Folgende: Technisch versierte Menschen erschaffen sich ihre Gottheit selbst, sie bauen sie mit und als Künstliche Intelligenz, in der Überzeugung, dass dies technisch ja in greifbarer Nähe sei und auch gut funktionieren werde. Diese mächtige KI-Gottheit könne man dann anbeten und ihr gleichzeitig aber auch klar machen, dass sie nicht einfach so vom Himmel gefallen oder aus dem Nichts entstanden sei, sondern dass die Kirchengründer sie schließlich erschaffen hätten und dementsprechende Erkenntlichkeiten erwartet würden. Lange gehalten hat sie nicht, die ganze Sache hat nicht geklappt. Nach fünf Jahren wurde die Kirche wieder geschlossen und das verbliebene Kirchenvermögen gespendet[2].

# 2   „Denkende" Maschinen und publikumswirksame Inszenierungen

Wie könnte es aber zukünftig um „denkende" Maschinen stehen? Das Bild der „Denkmaschine" macht es uns leicht, über KI-Systeme und ihre mitunter hochbeeindruckenden Ergebnisse zu sprechen. Allerdings ist es falsch. Da wird nichts gedacht. Wir nennen das bloß so.

Ein solches Denken, wie es jede und jeder von uns aus dem eigenen Selbsterleben kennt, findet im Digitalcomputer nicht statt. Im Digitalcomputer gibt es kein denkendes Subjekt, dort lebt nichts, was in diesem Verständnis denken könnte.

Für den Computerpionier Konrad Zuse waren denkende Maschinen schon im letzten Jahrhundert ein Thema. Er warnte angeblich davor, dass man bei Computern niemals den von ihm so genannten „letzten Draht" festlöten dürfe, denn wenn durch diesen letzten Draht Strom fließe, könnte – oder würde – im Computer ein Bewusstsein entstehen[3]. Zieht man die heutigen Rechenleistungen und Übertragungsgeschwindigkeiten und den Fortschritt bei KI-Systemen in den letzten Jahre in Erwägung, könnte man sich schon vorstellen, dass da noch einiges mehr drinsteckt. Und mal unter uns: Würden uns denkende Maschinen nicht von vielerlei lästigem Selberdenken-müssen entlasten? Wäre es nicht angenehm, wenn die unangenehmen Entscheidungen an eine denkende Maschine abgetreten werden könnten, delegiert an eine dem Menschen kognitiv ebenbürtige „starke KI" oder gar eine Superintelligenz? Könnten wir unser Denken nicht einfach „outsourcen" und in der Cloud erledigen lassen? In mancherlei Hinsicht gewiss ein verführerischer Gedanke.

Möglicherweise haben Sie vom Roboter „Sophia" gehört. Sophia ist eine dialogfähigen KI in Frauengestalt. Sophia hielt 2017 einen Konferenzvortrag in Riyadh in Saudi Arabien – und wurde vom Fleck weg als Staatsbürgerin in das Königreich eingebürgert. Manchen Bestandsbürgerinnen und Bürgern ging dies allerdings etwas zu schnell – und sie stellten Rückfragen der Art: „Wie kann es sein, dass diese KI nun mehr Bürgerrechte hat als jene Söhne unserer Frauen, deren Vater kein Saudi ist"? Und: „Hat dieser Roboter überhaupt die richtige Religion?"[4]. Diese Menschen hatten offenbar den Mut, sich ihres eigenen Verstandes zu bedienen und sich nicht von der technischen Inszenierung einlullen zu lassen.

Womit haben wir es also zu tun? Sind und bleiben „denkende Maschinen" ein Wunschtraum, wären sie eine Bedrohung (wenn es sie denn gäbe) oder sind sie ein schlichter Zuschreibungsfehler (zu wenig) denkender Menschen, weil es sich bloß um unbelebte Materie handelt und letztlich kein Subjekt, kein Leben in ihnen steckt?

## 3   Zukunftsszenarien

Angenommen, es gäbe sie irgendwann, die „denkenden" Maschinen. In den vergangenen Jahren bin ich solchen Behauptungen etwas intensiver nachgegangen, dass Künstliche Intelligenz irgendwann ein Bewusstsein haben könnte. In aller Kürze lässt sich zusammenfassen, dass fast niemand der weltweit hierzu Forschenden den aktuellen Digitalcomputern zutraut, ein phänomenales Bewusstsein zu entwickeln – also Bewusstsein in dem Sinne, dass die Maschine sich selbst als existierend erleben würde.

Aber was könnte die Zukunft bringen? Sollten Neurowissenschaftler irgendwann im Gehirn ein neuronales Korrelat des Bewusstseins finden – wonach manche eifrig suchen – wäre die Idee nicht weit, dies nachzubilden, z.B. mit neuromorphen Computern, und auszuprobieren, was passiert. Vielleicht wären die Ergebnisse auf alle Zeit enttäuschend – und nichts käme dabei heraus. Und wenn doch? Möglicherweise müssten diese KI-Maschinen auch bei fast Null anfangen, jeweils laufen lernen, sprechen lernen, lernen wie das sensorisch Gesehene, Gehörte und Getastete, vielleicht auch Gerochene und Geschmeckte zu Wahrnehmungseinheiten integriert wird. Vielleicht müssten diese Maschinen auch irgendwann durch die Pubertät, und müssten lernen, dass in anderen Gruppen andere Regeln gelten und Gepflogenheiten sehr unterschiedlich ankommen können.

Wir sprechen gerade über zukünftige Szenarien, die eintreten können, aber nicht müssen. In der Technikfolgenabschätzung hat man es hiermit öfters zu tun und versucht, zu sortieren und Eingriffsmöglichkeiten und Handlungsfelder aufzuzeigen. Wenn wir es also dereinst mit denkenden Maschinen zu tun hätten – wer sollte dann darüber entscheiden ob – und nun kommt der Vortragstitel wieder ins Spiel – diese Maschinen denken können *dürfen?* Wer könnte dies erlauben oder verbieten? Wer hat eigentlich das Mandat, einer denkenden Entität das Denken einzuschränken, zu verbieten oder zuzugestehen? Wer darf die Entitäten dumm halten?

In unserem Kulturkreis denken wir bei solchen Fragen schnell an Gesetze, Regulierungen und normative Vorgaben. Was aber, wenn die Regulierer mangels Gesamtverständnis unbrauchbare Vorgaben festsetzen? Ich sehe darin ein echtes Problem. Was wäre – und mit diesem Szenario habe ich mich etwas ausgiebiger beschäftigt – wenn das ganze Denken der Maschinen nur eine nahezu perfekte Imitation unserer Verhaltensweisen ist? Wenn diese Maschinen also gewissermaßen professionell inszenierte „Zombies" wären, so überzeugend, dass Regulierer, Lawmaker und viele andere ihnen auf den Leim gehen? Wenn plötzlich zivilgesellschaftliche Gruppen sich für Roboterrechte starkmachen, und dabei nur guten Imitationen aufsitzen, an ihre Sache aber fest glauben?

Als direkten Kontrast hierzu möchte ich Ihnen in einem Satz noch ein zweites Szenario gegenüberstellen: Möglicherweise entstehen dereinst tatsächlich denkende Maschinen, mit Bewusstsein, ohne dass wir dies in unseren ab- und aufgeklärten Betrachtungsweisen erkennen würden. Wir würden bewusste Entitäten lediglich als gut gelungene Imitationen ansehen – mit etlichen problematischen Konsequenzen für solche bewussten Maschinen.

## 4   Selbstaufklärung der technischen Disziplinen (statt noch einer Ethik)

Bevor ich gleich mit einer kleinen Abschlussgeschichte schließe, möchte ich die Notwendigkeit von auf diese Spezialthemen ausgerichteter Ethik betonen. Ethik braucht man ja dann, wenn sich aus der etablierten Moral keine passenden Antworten ableiten lassen, wenn neue Situationen neu zu bewerten sind. Die Entwicklungen der KI in diesem besonderen Feld erfordern mindestens ein ethisches Monitoring, besser noch ein aktives Mitgestalten der Entwicklungen durch gut ausgebildete Ethiker aus verschiedenen Fachdisziplinen, Leute, die auch die Technik verstehen. Die Aufgabe besteht darin, zu solch anstehenden Fragen, von denen ich heute ein paar skizzieren konnte, Orientierung zu beschaffen, Fantasien zu entmystifizieren, neue Herausforderungen zu objektivieren und besprechbar zu machen. Dafür werden sie gebraucht. Ich persönlich stehe für eine Stärkung der Selbstaufklärung der technischen Disziplinen, in denen gute Technikerinnen und Techniker noch besser werden, indem sie selbst technikethische Kompetenz aufbauen – und damit zu reflektierten Technikgestaltern werden können.

## 5  Politiker durch künstliche Intelligenz ersetzen?

Zum Abschluss die versprochen kleine Geschichte, die an der provokativen Frage ansetzt, ob man nicht auch den gesamten politischen Betrieb durch künstliche Intelligenz ersetzen könnte.

In einem Vorort von Tokyo fand vor nicht allzu langer Zeit eine Bürgermeisterwahl statt. Auf einem Wahlplakat war tatsächlich ein KI-Roboter mit weiblichem Antlitz abgebildet. Und in der Tat belegte Matsuda Michihito – so der Name auf den Plakaten – bei dieser Wahl den dritten Platz. Die Story wurde von internationalen Medien aufgegriffen, und Wahlberechtigte in verschiedenen Ländern wurden gefragt, warum sie denn einer KI ihre Stimme geben würden. Die Antworten gingen in ähnliche Richtungen: KI entscheide neutral und rationaler als Menschen, sie sei gerechter, weniger anfällig für Eigeninteressen, für Korruption und Klientelpolitik, und sie verlange keine Pensionszahlungen im Alter. Dringt man tiefer in die Geschichte ein, stößt man – wenig überraschend – auf einen echten Menschen hinter dem KI-System, der selbstverständliche eine politische Agenda hatte. Wichtige Entscheidungen, so sagte der leibliche Matsuda Michihito, wolle er als Bürgermeister die Künstliche Intelligenz treffen lassen[5].

## 6  Diskussionseinstiege

Über KI halten wir uns in vielerlei Hinsicht selbst den Spiegel vor. Die Aufgabe ist es, Technik so zu gestalten, so dass sie uns nützt – und nicht, dass wir uns der Technik fügen oder unterwerfen. Ob Maschinen dereinst denken können werden oder nicht, das können wir heute nicht beantworten – und müssen dies auch aushalten. Ob man solch ein Denken verbieten oder regulieren können sollte, das können wir allerdings schon heute diskutieren, jetzt, im Plenum und mit Ihnen im Publikum. Auf diesen Austausch freue ich mich schon jetzt!

______________________________________

Translation of the speech manuscript

# Are machines allowed to (be able to) think?

## Why Artificial Intelligence Needs Ethics

## Karsten Wendland

### Belief in Artificial Intelligence

What on God's earth has the topic of "Artificial Intelligence" got to do with a Catholic Day? My suggested answer: A lot, because the new belief of many people in artificial intelligence, especially the uninformed naive belief, has more and more devotees and followers. Artificial intelligence is expected to do a lot, even that it might someday stand as a superintelligence above all that has existed so far, first to replace us humanity in our role as world shapers and then to eliminate us from it – whether as a punishment for our shameful behavior, whether as logical-rational conclusion in averting of dangers against the own machine existence (we could want to pull the plug). Or simply because the probability that it would run altogether better without mankind than with all of us, would be simply calculated as higher – and we could at least understand and at best also agree why it should be over with us. So far a few dark popular dystopias sketched with few strokes.

In California a few years ago a church was founded. It was the "Way of the Future Church", an AI church whose deity was to become an artificial intelligence[1]. The founding idea of Levandowski, a multimillionaire, was this: Technically savvy people create their own deity, they build it with and as an Artificial Intelligence, convinced that this is technically within reach and that it will work well. This powerful AI deity can then be worshipped and at the same time it can be made clear that it did not just fall from the sky or come into being out of nothing, but that the founders of the church ultimately created it and that corresponding gratuities are expected. It did not last long, the whole thing did not work out. After five years the church was closed again and the remaining church property was donated[2].

## "Thinking" machines and effective staging for the public

But what might the future look like with thinking machines? The image of the "thinking machine" makes it easy for us to talk about AI systems and their sometimes highly impressive results. However, the image is wrong. Nothing is thought. We just call it that.

Such thinking, as each and every one of us knows from our own self-experience, does not take place in the digital computer. There is no thinking subject in the digital computer, nothing lives there that could think in this understanding.

For the computer pioneer Konrad Zuse, thinking machines were already an issue in the last century. He reportedly warned that computers should never be soldered to what he called the "last wire", because if current flowed through this last wire, a consciousness could – or would – arise in the computer[3]. Considering today's computing power and transmission speeds and the progress in AI systems in recent years, one could imagine that there is a lot more to it. And just between us: Wouldn't thinking machines relieve us of a lot of tedious self-thinking? Wouldn't it be pleasant if unpleasant decisions could be ceded to a thinking machine, delegated to a "strong AI" or even a superintelligence that is cognitively equal to humans? Couldn't we just "outsource" our thinking and have it done in the cloud? Certainly a tempting thought in some respects.

You may have heard of the robot "Sophia." Sophia is a conversational AI in female form. Sophia gave a conference talk in Riyadh, Saudi Arabia, in 2017 – and was naturalized as a citizen of the Kingdom on the spot. For some existing citizens, however, this was a bit too quick – and they asked questions like, "How can it be that this AI now has more civil rights than those sons of our women whose father is not Saudi?" And, "Does this robot even have the right religion?"[4]. These people obviously had the courage to use their own reason and not be lulled by the technical spectacle.

So what are we dealing with? Are and remain "thinking machines" a pipe dream, would they be a threat (if they existed) or are they a simple attribution error of (too few) thinking people, because they are merely inanimate matter and ultimately there is no subject, no life in them?

## Future Scenarios

Let's assume that they existed at some point, the "thinking" machines. In recent years, I have been following more closely such claims that artificial intelligence might one day have consciousness. In a nutshell, it can be summarized that almost no one of the world's researchers on this matter trusts the current digital computers to develop phenomenal consciousness – that is, consciousness in the sense that the machine would experience itself as existing.

But what might the future hold? Should neuroscientists eventually find a neural correlate of consciousness in the brain – which some are eagerly seeking – the idea would not be far off to replicate this, for example with neuromorphic computers, and try out what happens. Perhaps the results would be disappointing for all time - and nothing would come of it. And if it does? Perhaps these AI machines would also have to start from almost zero, each learning to walk, learning to talk, learning how to integrate what is sensory-seen, heard, and touched, perhaps also smelled and tasted-into units of perception. Perhaps these machines would also have to go through puberty at some point, and would have to learn that different rules apply in other groups and habits can be received very differently.

We are just talking about future scenarios that may or may not occur. In technology assessment, we often have to deal with this and try to assort and point out possibilities for intervention and fields of action. So if one day we had to deal with thinking machines: Who should decide whether – and now the title of the talk comes into question again – these machines are allowed *to be able to* think? Who could allow or forbid this? Who actually has the mandate to restrict, forbid or allow a thinking entity to think? Who is allowed to keep the entities stupid?

In our culture, we quickly think of laws, regulations, and normative directivs when we think of such questions. But what if the regulators set useless defaults for lack of overall understanding? I see this as a real problem. What if the whole thinking of machines is just a near-perfect imitation of our behaviors? So if these machines were, in a sense, professionally staged "zombies," so convincing that regulators, lawmakers, and many others fell for them? What if civil society groups suddenly started advocating for the rights of robots, which are just good imitations, but the activists firmly believe in their mission?

As a direct contrast to this, I would like to give you a second scenario in one sentence: Maybe one day actually thinking machines will emerge, with consciousness, without us recognizing this in our clarified, rationalized ways of looking at everything. We would regard conscious entities merely as well-done imitations – which would have significant problematic consequences for them.

## Self-enlightenment of the technical disciplines (instead of yet another ethics)

Before I finish with a little closing story, I want to emphasize the need for ethics focused on these special topics. Ethics is needed, after all, when no suitable answers can be derived from established morality, when new situations have to be assessed. The developments of AI in this special field require at least an ethical monitoring, better still an active co-design of the developments by well-trained ethicists from different disciplines, people who also understand the technology. The task is to provide orientation on such issues, to demystify fantasies, to objectify new challenges and to make them discussable. That is what they are needed for. I personally stand for a strengthening of the self-enlightenment of the technical disciplines, in which professional technicians become even better by building up techno-ethical competence themselves – and can thus become reflected technology designers.

## Replacing politicians with artificial intelligence?

Finally, the promised little story that starts with the provocative question of whether the entire political establishment could be replaced by artificial intelligence.

Not long ago, a mayoral election was held in a suburb of Tokyo. Election posters actually featured an AI robot with a female face. And indeed, Matsuda Michihito – the name on the posters – came in third in that election. The story was picked up by international media, and voters in various countries were asked why they would vote for an AI. The answers were along similar lines: AI would be neutral and more rational than humans, would be fairer, less prone to self-interest, corruption and patronage, and it would not require pension payments in old age. Delving deeper into the story, one encounters, unsurprisingly, a real human being behind the AI system who, of course, had a political agenda. Important decisions, said the fleshly Matsuda Michihito, he wants to let, as mayor, the artificial intelligence make[5].

## Discussion Entries

Through AI, we are holding up a mirror to ourselves in many ways. The task is to design technology so that it benefits us - and not so that we bow or surrender to technology. Whether machines will be able to think one day or not is something we cannot answer today – and we have to endure it. However, we can already discuss today, in the plenary session and with you in the audience, whether we should be able to prohibit or regulate such thinking. I am already looking forward to this exchange!

---

[1] Wired.com (2017). Inside the First Church of Artificial Intelligence. The engineer at the heart of the Uber/Waymo lawsuit is serious about his AI religion. Welcome to Anthony Levandowski's Way of the Future. https://www.wired.com/story/anthony-levandowski-artificial-intelligence-religion/ (last visited on 25.05.2022)

[2] TheCrunch (2021) Anthony Levandowski closes his Church of AI. https://techcrunch.com/2021/02/18/anthony-levandowski-closes-his-church-of-ai/ (last visited on 25.05.2022)

[3] Wendland, K., & Vater, C. (2020). Die Gründerväter der KI machten sich über Bewusstsein keine Gedanken. Im Gespräch mit Christian Vater. Podcast Selbstbewusste KI, Folge 4, veröffentlicht am 6.10.2020. DOI: https://doi.org/10.5445/IR/1000124235

[4] Arab News (2017). Saudi Arabia becomes first country to grant citizenship to a robot. https://www.arabnews.com/node/1183166/saudi-arabia (last visited on 25.05.2022)

[5] Költzsch, T. (2018) Eine künstliche Intelligenz als Bürgermeister. https://www.golem.de/news/ki-kandidat-eine-kuenstliche-intelligenz-als-buergermeister-1804-133830.html (last visited on 25.05.2022)